\documentclass[fleqn,10pt,a4paper]{wlscirep}

\usepackage{lineno}
\usepackage{lipsum}
\usepackage[utf8]{inputenc}
\usepackage[T1]{fontenc}

\title{Non-Radiating Resonances: Anapoles Enabling Highly-Efficient Plasma Jets within Dielectric Structures}

\author[1*]{Muhammad Rizwan Akram}
\author{Abbas Semnani}

\affil{Department of Electrical Engineering and Computer Science, The University of Toledo, Toledo, Ohio 43606, USA}

\affil[*]{muhammadrizwan.akram@utoledo.edu}


\begin{abstract}

Plasma plays a pivotal role in numerous applications spanning the fields of medicine, industry, agriculture, and space exploration. When plasma interacts with air, it initiates unique chemical reactions, resulting in the creation of rare and highly sought-after reactive species. Typically, the generation of plasma jets for interaction with air relies on resonant cavities to enhance plasma efficiency. In this study, we have harnessed the innovative concept of non-radiating sources, known as anapoles, which utilize the lowest order multipoles—specifically, electric-electric dipole interactions—within a hybrid metallo-dielectric structure. This approach enhances the near electric field, facilitating gas breakdown for the realization of a plasma jet. The achievement of a dielectric plasma jet is remarkable in its own right, particularly when considering the open structure employed, which enables frequency tuning. Furthermore, the prototype we have demonstrated surpasses existing plasma jet technologies in several key aspects, including compactness, compatibility with planar fabrication techniques, power efficiency, cost-effectiveness, tunability, and a twofold increase in electron density compared to the highest levels achieved to date. With these substantial enhancements, our proposed device is poised to revolutionize the landscape of plasma source technology and open up exciting avenues for exploring novel applications.

\end{abstract}
\begin{document}

\flushbottom
\maketitle
\thispagestyle{empty}

\section*{Introduction}

The confinement of electromagnetic (EM) waves or light\cite{a1hsu2016,aa1luk2010,ab1st2010} within meta-structures on a sub-wavelength scale has emerged as a crucial platform for exploring fundamental wave-matter interactions\cite{a2mon2014,a3hsu2013,a4sil2014,a5mon2018} and a wide range of applications, particularly in wireless\cite{a8zan2021} and optical domains\cite{a6alu2008,a7alu2010}. In Wireless applications such as sensing\cite{a9zha2022}, filtering\cite{a10coh1968}, and energy transfer\cite{a8zan2021}, achieving high-quality wave confinement within resonating structures is paramount. Consequently, novel concepts like "bound states in the continuum"\cite{a1hsu2016} and "anapoles"\cite{a12kae2010,a13fed2013} have garnered significant attention for enhancing wave confinement and providing high-Q resonances. However, due to the non-existence theorem\cite{a1hsu2016}, perfect bound states in a continuum cannot be realized in isolated structures. Conversely, anapoles\cite{a11yan2019} represent charge-current configurations that can be achieved through external excitation, inducing multipole moments within the structure that produce equal but out-of-phase far-field radiation. 

The initial demonstration of anapoles harnessed the interference between electric and magnetic dipoles, primarily focusing on toroidal dipoles, as a means to minimize outgoing scattering. Subsequently, anapoles have evolved beyond the confines of toroidal dipoles and have been realized through various combinations of other multipole configurations\cite{a14zur2019,a15she2020}. This versatility has prompted extensive exploration within the context of plane wave scenarios\cite{a16zen2017,a17can2021,a18gur2019,a19mir2015,a20kap2020,a21luk2017,a22luk2017,a23par2020}. More recently, there has been a concerted effort to achieve dipole-like excitation of anapoles, a development that holds promise for more compact devices\cite{a24nem2017,a25zan2021,a8zan2021}. However, a significant challenge lies in the requirement for high dielectric constant materials, which has posed a substantial barrier to practical applications. In our prior work\cite{a26akr2023}, we successfully demonstrated a compact anapole technology utilizing the lowest-order electric dipoles in conjunction with readily available, low-dielectric-constant materials. This breakthrough opens the door to  developing wireless applications that leverage anapole technology.

While anapole sources are energy storage devices, their storage capacity is inherently constrained by the materials used in their construction. This limitation prompts us to consider the potential applications of anapoles. In the context of our work, we have envisioned their application in developing efficient plasma sources. The augmented electric field generated by the presence of anapoles has the potential to initiate gas breakdown at much lower powers than usual, enabling the creation of efficient plasma sources. Plasma, known as the fourth state of matter, plays a pivotal role in diverse fields, including medicine, materials processing, agriculture, and space applications. In medicine, for instance, cold atmospheric plasma finds application in wound healing, cancer treatment, and dental procedures, thanks to the generation of many reactive oxygen and nitrogen species\cite{a27lar2015,a28wel2016}. Material processing leverages plasma for surface treatment, coatings, welding, and the processing of quartz and ceramics\cite{a29pen2015,a30sha2002}. In agriculture, plasma technology is extensively explored for pre-planting, pre-harvest, and post-harvest applications, including DNA modification, disinfection, and pathogen control, among others\cite{a31ito2018}. Furthermore, plasma holds importance in electric propulsion\cite{a32raf2021} and fusion technologies\cite{a33bet2016}. Given its profound impact across various aspects of daily life, developing efficient plasma sources is of paramount importance.

Numerous approaches have been investigated for plasma ignition, including those based on DC\cite{a34moh2002}, pulse\cite{a35wal2007}, and RF/microwave sources\cite{a36par2000, a37sch2008}. Among these, high-frequency-driven plasmas have garnered attention due to their operation in the $\alpha$-discharge regime \cite{a38rai2017,a39lie1994,a40sem2016,a41sem2016}, characterized by a low sheath voltage, which complements plasma discharge stability while addressing the significant issue of electrode erosion that affects the device's lifespan. High-frequency plasmas employ two main approaches. Non-resonant sources\cite{a42bar2010}, while effective, tend to be bulky and demand substantial power, giving rise to safety and electromagnetic interference concerns. In contrast, resonant plasma sources (RPS) offer an appealing alternative. In RPS, electromagnetic energy is concentrated at specific spatial locations within resonant structures, creating favorable conditions for gas breakdown and plasma formation. These RPS exhibit a range of advantageous features, including stable discharges, a higher degree of ionization and dissociation, elevated electron density, increased production of reactive species, lower plasma discharge temperatures, and reduced ignition voltage. Various solutions, such as the surfatron\cite{a43moi2022}, slot antennas\cite{a44kor1996}, cavity resonators\cite{a45gul2015}, and microstrip split ring resonators\cite{a46iza2003} have been proposed as resonant plasma sources, showcasing the versatility of this approach. 

Recently, an atmospheric pressure microwave plasma jet was developed, showcasing the ability to operate at a significantly reduced power of 500 mW, utilizing evanescent mode cavity resonator technology\cite{a47sem2022}. Similarly, plasma jets based on coaxial transmission line resonators\cite{a48cho2010,a49cho2009} have demonstrated operation at remarkably low input power levels of 1.5 - 3 W. Furthermore, to simplify and reduce the cost of manufacturing these intricate 3D structures, a novel approach involving substrate-integrated-waveguide (SIW) based cavity resonators\cite{a50zha2023} was introduced. This innovation enables the creation of a printed circuit board (PCB)-compatible plasma jet operating within the range of 15-30 W. In parallel, SIW-based evanescent cavity resonators\cite{a51kab2023} have been employed to achieve plasma jets operating at even lower power levels, typically ranging from 2.7 to 5 W.

A common issue with these resonating plasma sources is the impedance mismatch before and after plasma ignition. This mismatch can reflect a high-power signal, posing a risk to the generator system. The severity of this problem escalates as the input power requirements increase, mainly when aiming for higher gas flow rates and longer jet lengths\cite{a47sem2022, a48cho2010}. Recently, a novel physical concept known as "virtual perfect absorption" (VPA)\cite{a52del2023} has been introduced to mitigate these reflection issues. VPA involves the excitation of an exponentially increasing complex input signal, which prevents the impedance mismatch by adjusting the incoming power concerning the internal losses and reflected power. However, the practical application of VPA remains limited and has, thus far, been applied primarily under controlled conditions.

This study introduces a novel approach for developing highly efficient atmospheric plasma jets with minimal reflection and radiation losses using dielectric anapole structures. Unlike commonly employed cavity resonators, anapoles do not necessitate a metallic enclosure for radiation mitigation. The anapole structure designed for the plasma jet application leverages only the lowest-order electric dipoles induced on both the metallic and dielectric components of the device to minimize radiation losses through the destructive interference of outgoing waves. The proposed device is planar, exceptionally compact, and
seamlessly integrates with printed circuit board structures. Its cost-effectiveness is owed to its compatibility with PCB fabrication processes. Furthermore, the resulting plasma jet exhibits remarkable absorption efficiency, reaching as high as 94\% at a low input power of 1.5 W and 62\% at a high input power of 15 W. It provides double the electron density compared to state-of-the-art resonant microwave plasma jets.

\section*{Anapole Plasma Jet Design}
The proposed anapole plasma jet is centered on the excitation of the lowest-order multipoles, i.e., electric dipoles, utilizing a hybrid metallo-dielectric structure. The induced electric dipoles on both the metallic and dielectric components are strategically aligned to effectively cancel out far-field radiation while simultaneously enhancing the concentration of the near field. This interaction results in a substantial increase in the electric field strength in the vicinity of the device, which is used for gas breakdown. The detailed design and analysis of the metallo-dielectric anapole structure, using Cartesian multipole analysis, is provided in our previous work\cite{a26akr2023}. Leveraging the insights gained, the anapole structure was optimized to realize an efficient atmospheric pressure plasma jet operating at 2.45 GHz. 

\begin{figure}[ht]
\centering
\includegraphics[width=\linewidth]{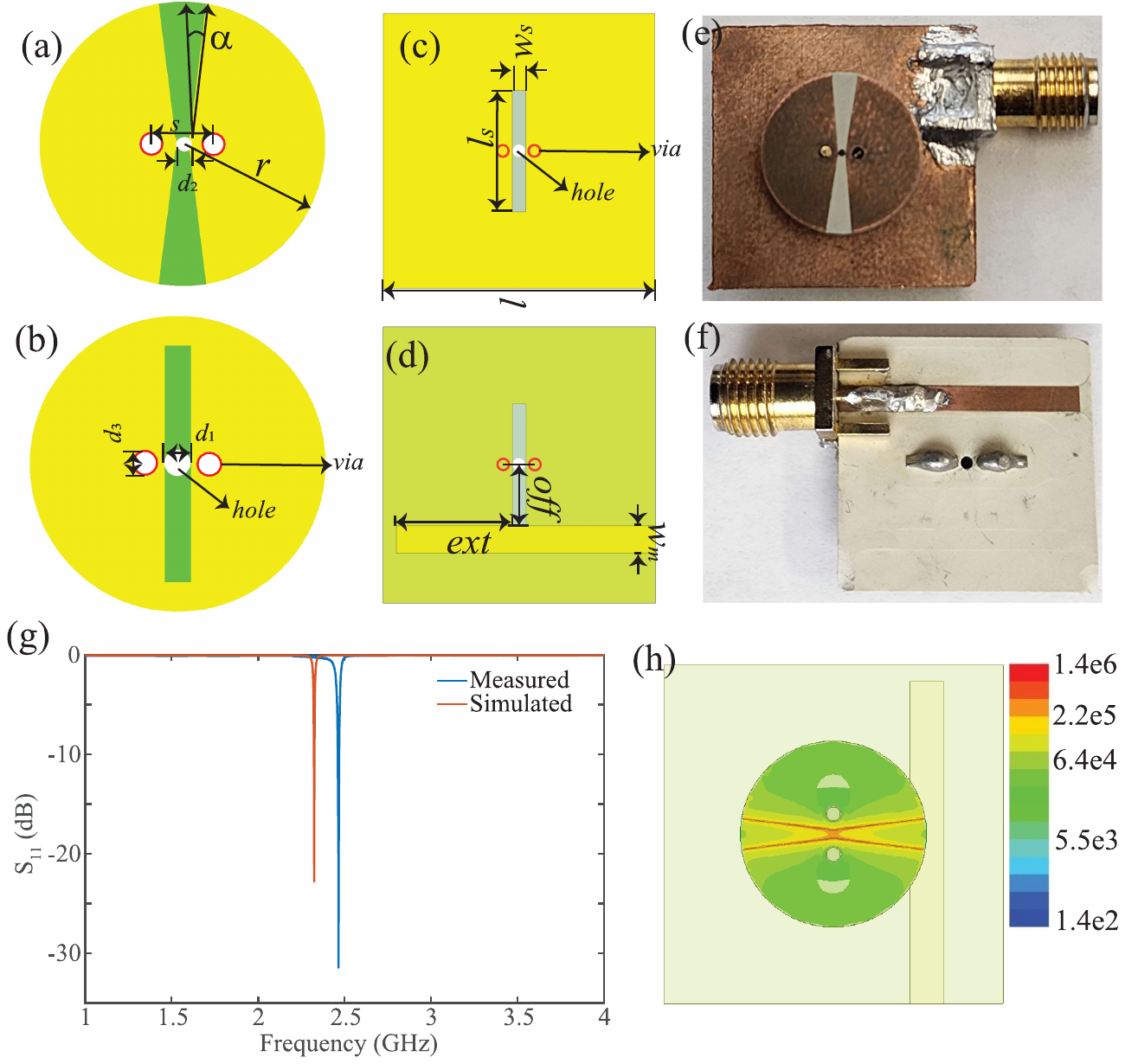}
\caption{Top (a,c) and bottom (b,d) views of the dielectric resonator and the feeding board of the anapole device with \textit{r} = 11 mm, thickness of the dielectric cylinder \textit{h} = 3.81 mm, thickness of feed board $h_2$ = 1.27 mm, $l_s$ = 8.83 mm, $l$ = 20 mm, $w_m$ = 2 mm, $ext$ = 9 mm, $off$ = 5.3 mm, $w_s$ = 1 mm, $d_1$ = 1 mm, $d_2$ = 0.5 mm, $d_3$ = 0.9 mm and $s$ = 1.2 mm. The cylindrical resonator is made of TMM13i with permittivity of $\epsilon_r$ = 13 and $tan\delta$ = $1.9\times10^{-3}$. The bottom board is made of TMM6 with permittivity of $\epsilon_r$ = 6 and $tan\delta$ = $2.3\times10^{-3}$. The fabricated anapole plasma jet device (e) top view (f) bottom view. (g) The simulated and measured reflection coefficients showcase a solid resonance and (h) Enhanced electric field in the range of $10^6$ (V/m) at the gas ignition region.}
\label{fig1:design}
\end{figure}

The designed anapole structure consists of two separate boards: a cylindrical dielectric disk and a feeding microstrip board. The cylindrical disk is intricately fashioned from a commercially available Rogers TMM13i laminate, possessing a thickness of 3.81 mm and a 35 $\mu$m copper cladding. This disk serves as the resonant element of the anapole plasma jet and incorporates vias to create a split-ring resonator configuration. The top and bottom copper patterns on the disk facilitate longer current paths, contributing to the compactness of the design. Additionally, a rectangular slot is etched into the disk's bottom layer to enable electromagnetic energy coupling from the feeding network to the dielectric resonator cavity. For the feeding board, another layer is fabricated from a 1.27-mm thick Rogers TMM6 laminate featuring the same 35 $\mu$m cladding with a 50-$\Omega$ microstrip line on the bottom side and a slot of identical dimensions on the top. The positioning of the microstrip line, slot width ($w_s$), and length ($l_s$) have been meticulously optimized to achieve excellent impedance matching at the resonant frequency of 2.45 GHz. The assembly process involves aligning the cylindrical disk with the feeding board using vias and metallic rods, which are then securely affixed using silver epoxy. The resulting prototype exhibits remarkable compactness, with its largest dimension measuring less than 0.1$\lambda$. Detailed design schematics and images of the fabricated device are provided in Fig.~\ref{fig1:design}(a-f).

To establish the necessary gas flow mechanism for plasma jet formation, a 1-mm hole is drilled between the rods and extends to the center of the cylindrical disk. As it approaches the disk's surface, the hole is tapered down to a diameter of 0.5 mm. This final hole size ultimately determines the diameter of the plasma jet. A Teflon capillary tube is then threaded through this hole, extending from the feeding board to the midpoint of the disk. This capillary tube serves as the conduit for gas injection. Positioning the hole at the disk's center within the dielectric resonator cavity is a deliberate choice, ensuring a sufficiently strong electric field for facilitating plasma ignition, as illustrated in Fig.~\ref{fig1:design}(h). Furthermore, the slot on the top side of the disk is tapered to minimize the risk of air breakdown at various points along its length.

\section*{Results and Discussions}

The frequency response of the proposed device was initially assessed in the OFF mode, employing both numerical simulations and experimental measurements, as illustrated in Fig. 1(g). Notably, the gas flow was found to have an insignificant impact on the resonance frequency and can thus be considered a non-significant factor in this context. In the experimental measurements, a slight shift of approximately 130 MHz in the resonance frequency is observed compared to the simulation results. This deviation can be attributed to the fabrication tolerances inherent to the manufacturing process, as the initial design had aimed for a resonance frequency of 2.45 GHz. Remarkably, the achieved return loss, surpassing 20 dB, underscores that over 99\% of the input microwave energy is efficiently coupled to the anapole device at the resonant frequency.

To induce gas breakdown, it is crucial to establish a high electric field of around $10^5$ (V/m) within the gas flow region. The electric field distribution was numerically evaluated in the proposed anapole design, utilizing an input power of 1 watt, as illustrated in Fig.~\ref{fig1:design}(h). Here, a maximum electric field strength of $1.4\times10^5$ (V/m) is observed, accompanied by a radiation loss of approximately 250 mW out of the 1-W input power. It is important to mention that a potential strategy for reducing radiation loss involves employing a narrower slot on top of the disk without tapering. However, such an approach might increase the risk of undesired air breakdown, which is undesirable in the context of a plasma jet device. Furthermore, it's noteworthy that the peak electric field strength remains unaffected by the heightened radiation loss resulting from slot tapering.

\begin{figure}[ht]
\centering
\includegraphics[width=\linewidth]{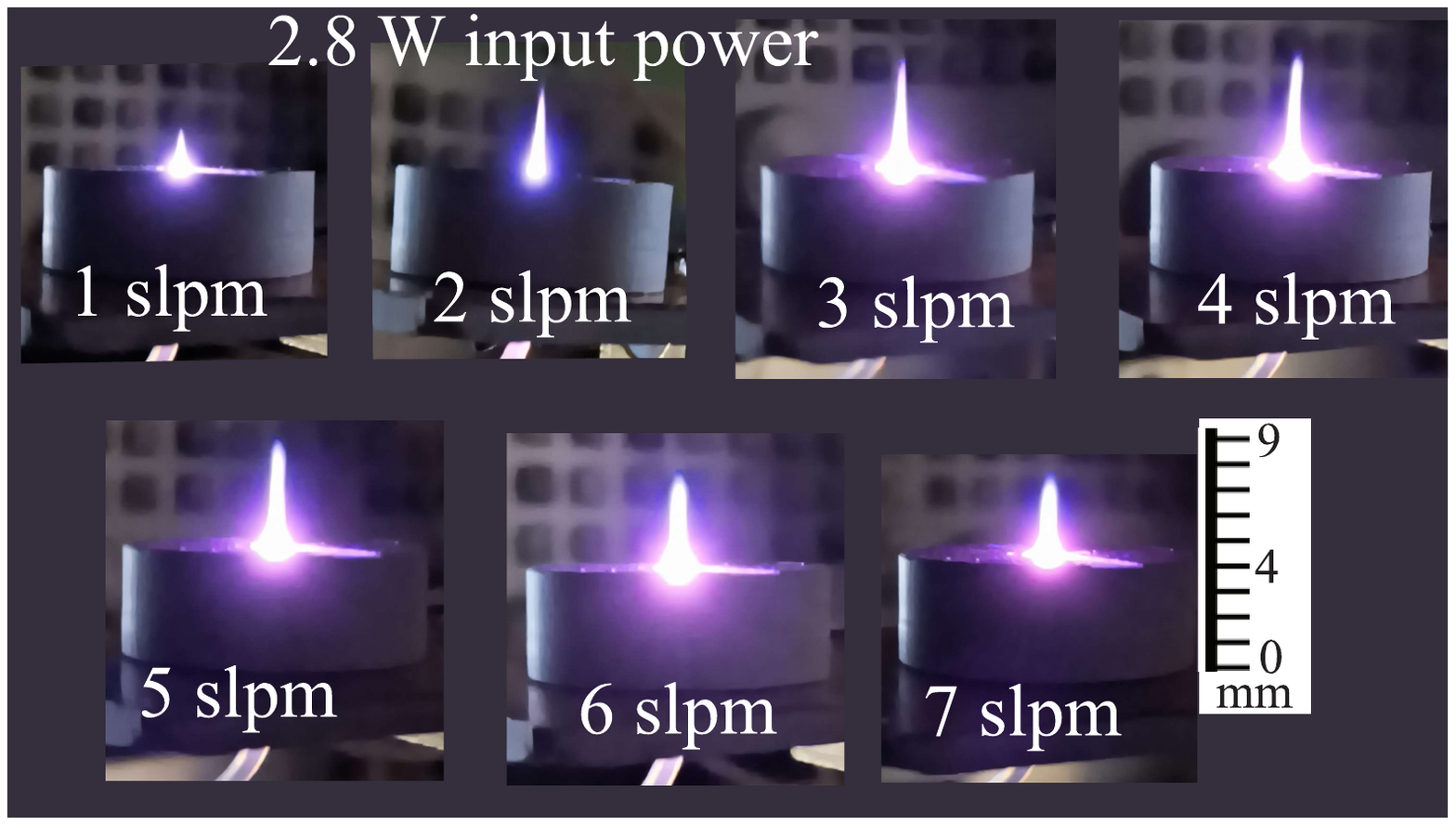}
\caption{Images of the anapole plasma jet at various helium flow rates at a constant input power of 2.7 W.}
\label{fig3:Jets}
\end{figure}

To test the device in ON mode, helium gas was introduced into the device at 1 slpm. At 2.7 W of input microwave power at the resonant frequency, gas breakdown occurred, leading to the formation of a plasma jet. Due to the reduced effective area post-breakdown, even lower power in the 1-W range proved sufficient to sustain the plasma jet after ignition. To gain further insight into the characteristics of the plasma jet, the helium flow rate was varied, ranging from 1 to 7 slpm, and the resulting plasma jets are depicted in Fig.~\ref{fig3:Jets}. It is noteworthy that the plasma jet length exhibited an increasing trend with increasing flow rate until it reached 5 slpm, at which point an optimal plasma jet configuration was observed. Beyond this flow rate, the jet length began to decrease. The nature of gas flow plays a crucial role in shaping the plasma jet and is contingent upon factors such as the channel type and gas properties. This relationship can be elucidated with the assistance of the Reynolds number. The Reynolds number is a dimensionless parameter that can be computed to gauge the degree of turbulence within the gas flow\cite{a53van2015} as

\begin{equation}\label{eqn1}
    Re = \rho vd/\eta,
\end{equation}

where $\rho, v, \eta$ represent the density, velocity, and viscosity of the fluid, respectively. In the context of gas flow within a cylindrical channel, $d$ is the channel diameter, which measures 0.5 mm in this study. A Reynolds number below 2000 signifies a laminar flow regime, resulting in a uniform, needle-like jet, while a value exceeding 3000 indicates turbulent flow, which is generally undesirable. The gas velocity can be calculated based on the gas flow rate under standard conditions as follows:

\begin{equation}\label{eqn2}
    v = 4D/\pi d^2,
\end{equation}

where D represents the gas flow rate in slpm. Helium possesses a density of $\rho = 0.1634$ kg/m$^{3}$ and a viscosity of $\eta = 1.94 \times 10^{-5}$ kg/(m.s). According to \ref{eqn1}, when D is less than 5 slpm, the calculated Reynolds number remains below 2000, indicating a laminar flow regime. This laminar flow is desirable for achieving a uniform plasma discharge, a characteristic that aligns well with the experimental observations presented in Fig.~\ref{fig3:Jets}.

\subsubsection*{Plasma Jet Efficiency}
\begin{figure}[ht]
\centering
\includegraphics[width=\linewidth]{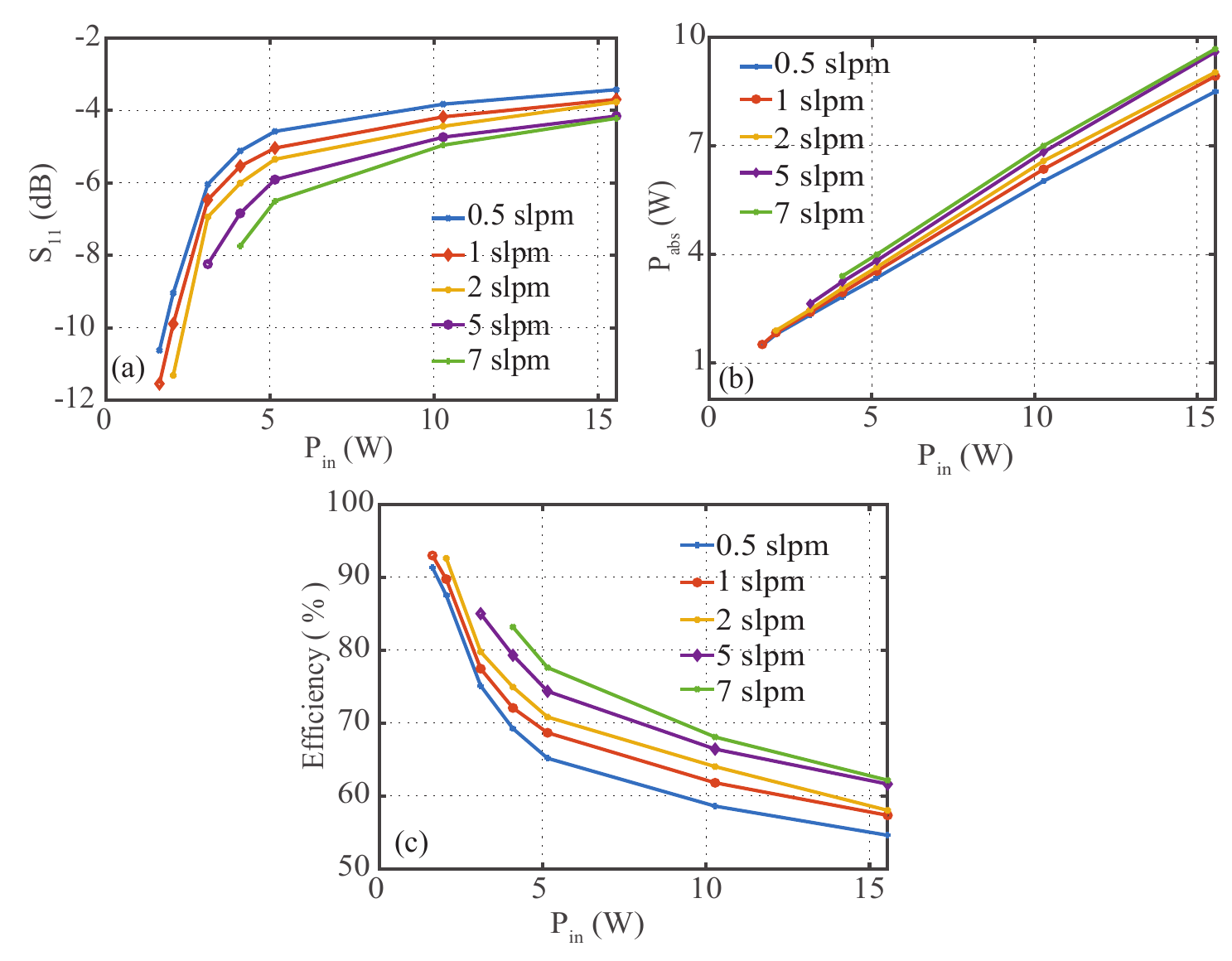}
\caption{Measured (a) reflection coefficients, (b) absorbed powers, and (c) power efficiencies versus input power at various flow rates.}
\label{fig4:efficiency}
\end{figure}

 The plasma jet's efficiency is experimentally characterized, as detailed in the "Efficiency Evaluation" section in "Materials and Methods" and \ref{fig2:setup}. The device exhibits a reflection coefficient of approximately -20 dB in the OFF mode. However, when plasma ignites, it perturbs the frequency response due to the creation of a relatively conductive plasma region positioned directly within the high-field region of the structure. Experimental measurements are conducted to determine the input and reflected powers at various flow rates, enabling the calculation of the absorbed power by the plasma. Based on these numbers, the device's efficiency is calculated and depicted in Fig.\ref{fig4:efficiency}. It is observed that the proposed anapole plasma jet consistently maintains a relatively low reflection, even at high operating powers. This characteristic highlights that the need for circulators and couplers can be circumvented without compromising the safety of the microwave sources, as is often required in the case of conventional resonant plasma jet sources. The absorption efficiency of the proposed device is impressively high, reaching 94\% at a low input power of 1.5 W and 62\% at a higher input power of 15 W. This level of efficiency sets the proposed anapole device apart from earlier plasma jets, particularly those designed for low-power operation. For instance, the efficiency of the anapole device significantly surpasses that of evanescent-mode cavity resonator-based plasma jet\cite{a47sem2022}, which achieved 80\% efficiency at 1 W and 18\% efficiency at 15 W input power. Similarly, the proposed anapole device outperforms coaxial transmission line resonator-based plasma jets\cite{a48cho2010,a49cho2009}, which attain 80\% efficiency at a low input power of 1.5 W.  

\subsubsection*{Plasma Jet Diagnostics}
\begin{figure}[ht]
\centering
\includegraphics[width=\linewidth]{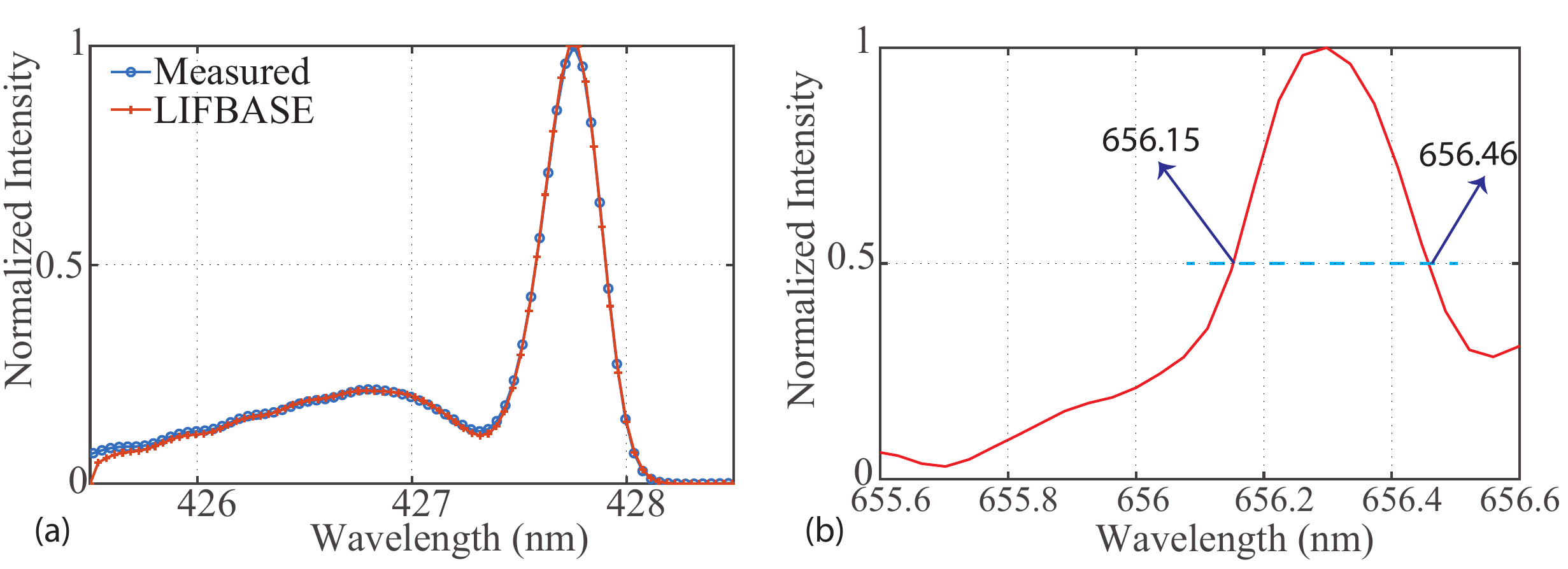}
\caption{(a) Excellent agreement between the simulated and measured spectrums at 5 slpm flow rate and 15 W input power for a temperature of 350 K. (b) H-$\alpha$ line at 15 W input power and 5 slpm of helium flow rate. $\Delta \lambda_{Stark}$ is calculated from FWHM as required
to extract the jet electron density.}
\label{fig5:diagnostics}
\end{figure}

Critical parameters governing the usability of a plasma jet for various applications include characteristics such as gas temperature and electron density. To ensure the safety of the plasma jet, it is essential to maintain a temperature close to room level. In this context, the plasma discharge temperature was characterized by evaluating the rotational gas temperature for diatomic molecules N$_2^+$ \cite{a54cha2012}. To achieve this, optical emission spectrometry (OES) was employed, utilizing a high-resolution optical sensor to capture the spectral profile of N2+ molecules over a 10-second duration (refer to the 'Spectral Evaluation' section in 'Materials and Methods'). Subsequently, the experimentally obtained profile was compared with the spectrum generated by LIFBASE for N$_2^+$ molecules. By comparing the two spectral profiles, the discharge temperature was accurately calculated with a $\pm$5 K precision. As an illustration, for an input power of 15 watts and a gas flow rate of 5 slpm, the experimental and simulated profiles are juxtaposed in \ref{fig5:diagnostics}(a), showcasing an excellent alignment between the two profiles at a temperature of 350 K. The gas temperature remains at 315 K for an input power of 5 W and rises to 350 K at an input power of 15 W. This temperature range ensures the device's safety for temperature-sensitive applications, such as plasma medicine.

Determining the gas temperature makes it possible to characterize the electron density of the anapole plasma jet. To achieve this, optical emission spectrometry is employed to analyze spectral profiles. Among the commonly utilized spectral profiles in this context are the spectral emissions of hydrogen atoms, specifically Balmer-alpha (H-$\alpha$) at 656.279 nm and Balmer-beta (H-$\beta$) at 486.135 nm, due to their visibility in the spectrum and distinct linear Stark effect. In this study, we opted to employ the H-$\alpha$ lines for estimating the electron density (n$_e$), as they offer greater distinctiveness when compared to the H-$\beta$ spectral profiles. A spectral profile within the atmospheric plasma jet can be described as a convolution of Gaussian and Lorentzian profiles, collectively referred to as the Voigt function. The Gaussian component of the spectral profile is influenced by factors such as the mass of the hydrogen atom, central wavelength, and gas temperature. Conversely, the Lorentzian component, which is more dominant, encompasses Doppler, van der Waals, and Stark broadening effects. It's important to note that resonance broadening due to interactions between neutral atoms of the same kind, while present, is typically negligible for Balmer lines at atmospheric pressure and can, therefore, be disregarded. The remaining three broadening mechanisms are accurately considered when assessing the electron density. Doppler broadening, for instance, arises when emitting atoms exhibit random motion, and the full width at half maximum can be determined as\cite{a55nik2015}:

\begin{equation}
    \Delta \lambda_D = \lambda_0(8ln2\dfrac{K_b T_g}{m_a c^2} ).
\end{equation}

Here, the gas temperature $T_g$ is in Kelvin, Boltzmann's constant $K_b$ is in $JK^{-1}$, and $m_a$ denotes the mass of the emitter. Van der Waals broadening, on the other hand, arises due to interactions between atoms of different species and can be estimated as\cite{a56wan2018}:

\begin{equation}
    \Delta \lambda_{vdW} = \dfrac{C}{T_g^{0.7}},
\end{equation}

where C is a gas constant equal to 2.42 for helium. The Doppler and van der Waals broadenings are calculated using the above formulas for a gas flow rate of 5 slpm under various input powers, as detailed in Tab.\ref{tab:par}, along with the corresponding measured temperatures. Furthermore, the full width at half maximum (FWHM) of the H-$\alpha$ spectral line is determined from \ref{fig5:diagnostics}(b). It is noteworthy that $\Delta \lambda_D$ and $\Delta \lambda_{vdW}$ are found to be very small when compared to the FWHM of H-$\alpha$, indicating that Stark broadening is the dominant profile. Consequently, the electron density of the plasma jet is estimated using Stark broadening as\cite{a57gig2003}:

\begin{equation}
    n_e = 10^{17} \times ({\Delta \lambda_{Stark}}/1.098)^{1.47135}.
\end{equation}

 Here, n$_e$ is in cm$^{-3}$, and $\lambda_{Stark}$ is in nm. The evaluated n$_e$ is approximately 1.55$\times10^{16}$ cm$^{-3}$, at least twice the values typically achieved in conventional plasma jets employing resonant cavity approaches. 

\begin{table}[ht]
\centering
\begin{tabular}{|l|l|l|l|}
\hline
\textbf{Input Power} & 5 W & 10 W & 15 W \\
\hline
$T_g (K)$ & 315 & 330 & 350 \\
\hline
$\Delta \lambda_{Stark} (nm)$ & 0.3 & 0.305 & 0.31 \\
\hline
$\Delta \lambda_D (nm)$ & 0.01025 & 0.0105 & 0.01081 \\
\hline
$\Delta \lambda_{vdW} (nm)$ & 0.04315 & 0.04177 & 0.04008 \\
\hline
\end{tabular}
\caption{\label{tab:par} Measured gas temperature and the corresponding spectral broadening data for the 2.45 GHz Anapole plasma jet at various input power levels while maintaining a constant helium flow rate of 5 slpm.}
\end{table}

\subsubsection*{Frequency Reconfigurability}
The interaction between a plasma jet and the surrounding ambient air holds significant potential for generating highly reactive species, including OH, NO, NO$_2$, O, and O$_3$, among others. These reactive species have garnered considerable interest for their applicability in medical and agricultural contexts. The production of these reactive species is intricately linked to the specific attributes of the plasma jet, such as the background gas composition, the rate of gas flow, and the method employed to generate the plasma jet. For example, as demonstrated in \cite{a58van}, microwave plasma jets can yield substantial quantities of desirable NO species. The frequency of operation plays a pivotal role in influencing the production of a diverse array of reactive species. However, it is essential to note that the characteristics associated with frequency remain relatively underexplored, mainly owing to the complex nature of microwave plasma jet technologies. Recent developments in low-power operated plasma jets \cite{a47sem2022, a51kab2023, a49cho2009} have significantly enhanced the accessibility of plasma sources for generating and utilizing reactive species within the microwave frequency range. However, these technologies are primarily built around cavities with electric fields confined within metallic enclosures, making frequency-tunable operations complex.

\begin{figure}[ht]
\centering
\includegraphics[width=\linewidth]{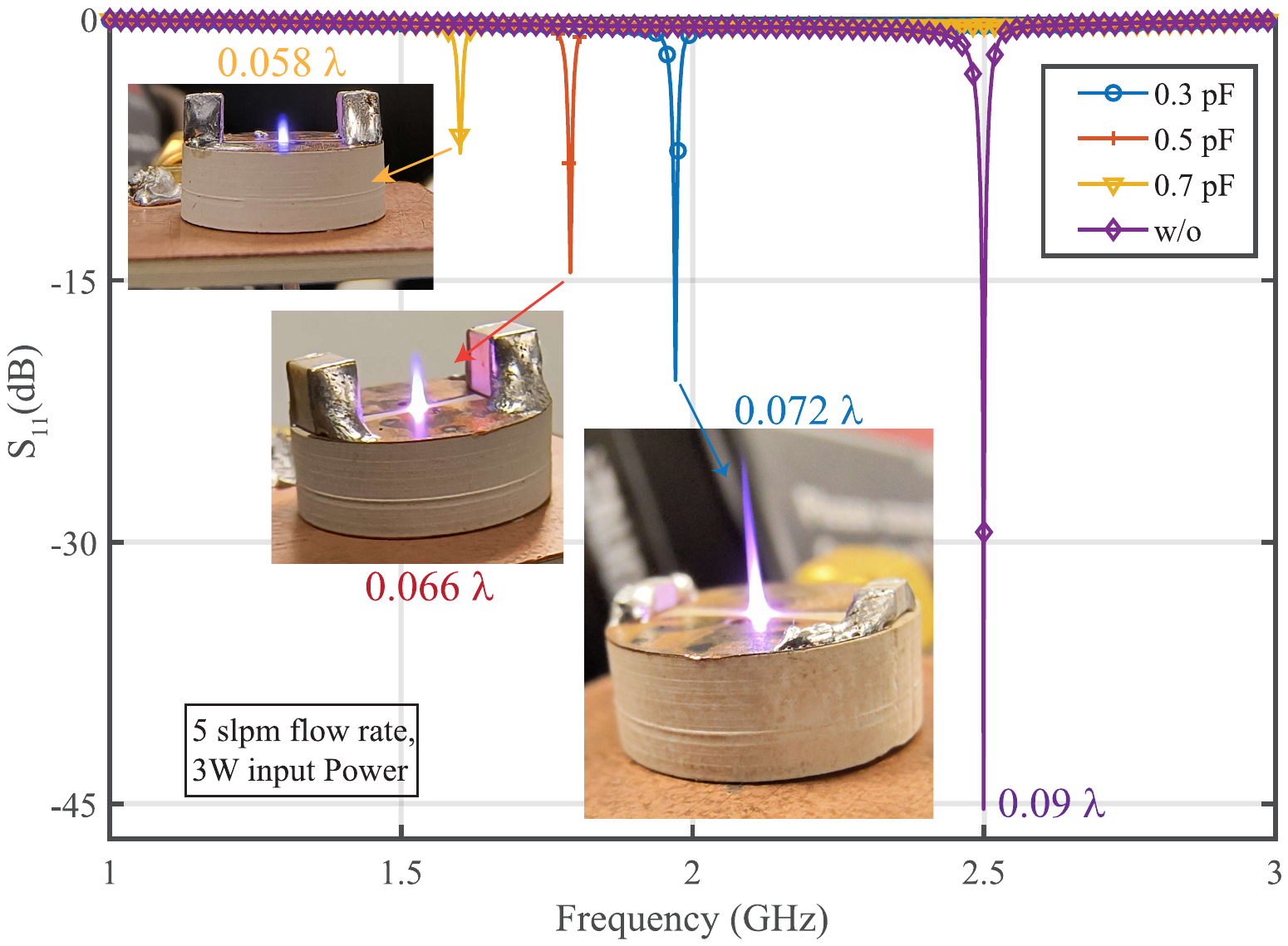}
\caption{The frequency tuning of the anapole plasma jet is realized by employing two tuning capacitors positioned at the edges of the tapered slots. By varying the capacitance within the range of 0.3 to 0.7 pF, the system achieves an impressive tuning capability, spanning nearly 900 MHz, with frequencies ranging from 2.5 GHz (no capacitor) to 1.6 GHz in the presence of the 0.7 pF capacitors.}
\label{fig6:reconfig}
\end{figure}

The presented anapole structure is designed as an open cavity featuring an accessible electric field for frequency tunability. As depicted in Fig.\ref{fig6:reconfig}, this tunability is achieved by using two capacitors positioned at the edges of a tapered slot on the dielectric disk. This design facilitates tunable operation within a broad frequency range, spanning from 1.6 GHz to 2.5 GHz, offering a 900 MHz bandwidth. To validate this concept, three prototypes were successfully demonstrated, each operating at different frequencies: 1.6 GHz, 1.8 GHz, and 1.97 GHz. In each case, a plasma jet was generated using a 5-slpm flow rate of Helium with only 3W of input power, highlighting the low-power operation of the system. Furthermore, adjusting the same design to a lower frequency of 1.6 GHz while maintaining the exact dimensions emphasizes its compactness, enabling scalability and facilitating seamless integration with microwave sources.

\section*{Conclusion}
This work successfully demonstrated a novel, fully planar, compact, and frequency-tunable atmospheric pressure plasma jet device. This plasma jet technology leverages the capabilities of a dielectric anapole structure, a non-radiating resonator, to enhance the near electric field while effectively suppressing far-field radiation. This demonstration represents a pivotal step in unlocking a new pathway toward highly efficient plasma sources characterized by minimal reflection and radiation. The key advantages of the proposed technology, including its high electron density, compact form factor, seamless integration capability, and cost-effectiveness, hold the potential to open up new horizons for discovery and application. Given the profound impact of plasma technology across many fields, these advanced attributes are positioned to make substantial contributions to progress in various domains. Moreover, the inherent ease of frequency tunability of the proposed technology holds the promise of facilitating explorations into enriched chemistry within the RF/Microwave spectrum, further enhancing its utility and versatility. 

\section*{Materials and Methods}
\subsubsection*{Simulation Methods}
The proposed anapole device is numerically evaluated using the High-Frequency Structure Simulator (HFSS) 2023 R1, employing the Eigenmode solver. This analysis yielded an estimated Q-factor reaching 256 at the design frequency of 2.45 GHz. Subsequently, the driven solution was used to assess the return loss near the resonance frequency, as illustrated in \ref{fig1:design}(g). Additionally, the radiation loss was determined, measuring 250 mW out of 1 W for the proposed plasma jet device.
\subsubsection*{Samples Preparations}
The proposed device consists of two distinct components: (1) a cylindrical disk and (2) a feeding board, both of which are amenable to printed circuit board (PCB) fabrication techniques. The cylindrical disk is crafted from Rogers TMM13i laminate and incorporates two vias, along with a central hole, to prevent unintended air breakdown. The feeding board is constructed using Rogers TMM6 material. It is affixed to the disk using metallic rods of smaller dimensions than the vias, ensuring proper slot alignment. Silver epoxy is employed to fasten the disk atop the feeding board securely. Microwave energy is coupled to the device via a 50-$\Omega$ SMA connector.  
\subsubsection*{Experimental setup}
\begin{figure}[t]
\centering
\includegraphics[width=\linewidth]{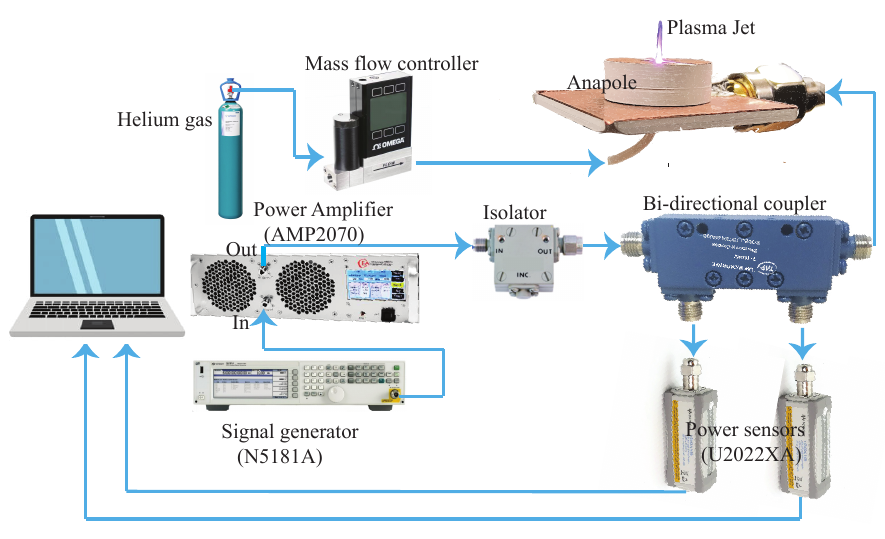}
\caption{Experimental setup for the anapole plasma jet ignition and characterization.}
\label{fig2:setup}
\end{figure}

After the sample preparation, a vector network analyzer (VNA) is employed to measure the scattering parameters, as depicted in \ref{fig1:design}(g). Given that the device is an open structure, it is essential to assess the stability of its frequency response to ensure optimal performance. A slight shift in the resonance frequency, around 1 MHz, is observed when a metallic sheet is brought into proximity to the device within a range of 2 to 4 mm. The low radiation characteristics of the proposed anapole device make it well-suited for plasma ignition without the need for an enclosed metallic cavity. The setup for achieving plasma ignition is illustrated in Fig.~\ref{fig2:setup}. The signal generator N5181A generates a continuous wave (CW) signal at the resonance frequency, which is subsequently amplified by the AMP2070 power amplifier, providing approximately 57 dB of gain. The amplified signal passes through an isolator to safeguard the amplifier from back-reflected signals. Subsequently, the signal is routed through a 30-dB bi-directional coupler, where two power sensors, namely U2022XA, are employed to measure the input and reflected powers. A compressed helium gas cylinder is connected to a mass flow controller (MFC), which regulates the gas flow rate directed to the device. For a gas flow rate of 1 slpm, plasma ignition is achieved at an input power of 2.7 W and can be sustained with an input power as low as 1 W.

\subsubsection*{Efficiency Evaluation}
The aforementioned setup is employed to assess the plasma efficiency. Given that we measure the input and reflected powers using a bi-directional coupler, it is necessary to account for the losses incurred by cables and components between the reflected port and the input port of the anapole device. These losses are determined through a VNA measurement and are factored in for precise power measurements at the input port of the anapole device. The absorbed power within the plasma can be readily derived from the reflected power, and the efficiency is subsequently calculated using the absorbed power and input power. 
\subsubsection*{Spectral Evaluation}
We employed the Teledyne Princeton Instruments HRS-500-SS spectrometer, which offers an optical resolution of 0.05 nm, to assess the spectral emissions. The optical sensor was positioned in close proximity to the device, approximately 1-2 mm away from the side, to prevent interference with the gas flow. This placement was optimized to capture the most comprehensive spectral profile of the desired emitter. To ensure data stability and minimize variations, the sensor's exposure time was set to 10 seconds, allowing adequate time for the data to stabilize.

\bibliography{sample}

\section*{Acknowledgements (not compulsory)}

This work was supported by the National Science Foundation (NSF) under Grant ECCS-2102100.

\section*{Author contributions statement}

Must include all authors, identified by initials, for example:
A.A. conceived the experiment(s),  A.A. and B.A. conducted the experiment(s), C.A. and D.A. analysed the results.  All authors reviewed the manuscript. 

\section*{Additional information}

To include, in this order: \textbf{Accession codes} (where applicable); \textbf{Competing interests} (mandatory statement). 

The corresponding author is responsible for submitting a \href{http://www.nature.com/srep/policies/index.html#competing}{competing interests statement} on behalf of all authors of the paper. This statement must be included in the submitted article file.

\end{document}